\documentclass[12pt,english]{article}
\usepackage{mathptmx}
\usepackage{amsthm}
\usepackage{amsmath}
\usepackage{amssymb}
\usepackage{hyperref}

\def\beq{\begin{equation}}
\def\eeq{\end{equation}}



\usepackage{babel}

\begin{document}

\title{On The Relativistic Classical Motion of a Radiating Spinning Particle
in a Magnetic Field}

\author{Arnab Kar\footnote{arnabkar@pas.rochester.edu} \ and S. G. Rajeev\footnote{Also at the Department of Mathematics}}

\maketitle
\begin{center}
Department of Physics and Astronomy\\ University of Rochester\\
Rochester NY 14627\\ USA
\end{center}

\begin{abstract}
We propose classical equations of motion for a charged particle with
magnetic moment, taking radiation reaction into account. This generalizes
the Landau-Lifshitz equations for the spinless case. In the special
case of spin-polarized motion in a constant magnetic field (synchrotron
motion) we verify that the particle does lose energy. Previous proposals
did not predict dissipation of energy and also suffered from runaway
solutions analogous to those of the Lorentz-Dirac equations of motion.
\end{abstract}

\section{Introduction}

What is the classical motion of an electron in a constant magnetic
field? Surprisingly, this question does not have a definitive theoretical answer
even now. The complications arise when the  radiation from the particle
is considered. Without it, the answer is quite elementary: a helical
path around the magnetic field\cite{EM} obtained by solving the Lorentz
equation of motion.\footnote{In units with $c=1$. $q$ and $m$ are the charge and mass respectively.
Also, $u^{\mu}=\frac{dx^{\mu}}{d\tau}$ is the four-velocity and $\tau$
is proper time. Note the constraint $\eta_{\mu\nu}u^{\mu}u^{\nu}=1$
and its derivative $\eta_{\mu\nu}u^{\mu}\frac{du^{\nu}}{d\tau}=0.$ %
}

\beq
\frac{du_{\mu}}{d\tau}=\frac{q}{m}F_{\mu\nu}u^{\nu}
\eeq
But we know that matters are not so simple in reality: an accelerated
charge radiates, and the radiation emitted by it exerts a force back
on it. If the energy is not replenished the path should be a spiral,
the radius of the circle decreasing with the energy. 

Determining the force exerted on a point particle by its own field
is fraught with conceptual difficulties: the force is infinite. Dirac\cite{Dirac38}
made a major step forward by showing that this divergence can be removed
by a renormalization of the mass. The resulting equation (the Lorentz-Dirac
equation of motion) \footnote{The last term includes a projection operator that ensures that the
constraint $\eta_{\mu\nu}u^{\mu}\frac{du^{\nu}}{d\tau}=0$ is satisfied.%
}

\beq
\frac{du_{\mu}}{d\tau}=\frac{q}{m}F_{\mu\nu}u^{\nu}+\epsilon\left[\eta_{\mu\nu}-u_{\mu}u_{\nu}\right]\frac{d^{2}u^{\nu}}{d\tau^{2}},\quad\epsilon=\frac{2}{3}\frac{q^{2}}{m}
\eeq 
is still not the correct answer. It predicts runaway solutions: even
in the absence of any external fields, a particle can accelerate away
to infinity with exponentially \textbf{growing} energy. Part of the
problem is that the radiative terms persist even in the absence of
external forces. Also, the equation involves the third derivative
of position; so it needs a new initial condition. Dirac suggested
that this be chosen such that the energy remains finite forever. However,
such solutions suffer from another problem: a violation of causality;
the particle would accelerate \textbf{before} the external field is
turned on. 

The textbook of Landau and Lifshitz\cite{LandauLifshitz} proposed
(without proof) a way out of this dilemna. Replace the derivative
of acceleration by its leading approximation in an expansion in powers
of $\epsilon$:

\beq
\frac{du_{\mu}}{d\tau}=\frac{q}{m}F_{\mu\nu}u^{\nu}+\epsilon\left[\eta_{\mu\nu}-u_{\mu}u_{\nu}\right]\frac{d}{d\tau}\left[\frac{q}{m}F_{\rho}^{\nu}u^{\rho}\right],\quad\epsilon=\frac{2}{3}\frac{q^{2}}{m}
\eeq

The radiative term now vanishes when the extrenal force is zero; and
the equation now involves only the second order derivative of position.
The free particle does not run away. Further analysis shows that the
orbits in a constant magnetic field \cite{Spohn} and in a Coulomb
field \cite{Rajeev} decay as expected. But why should we stop at
first order in the iteration in powers of $\epsilon$? Although it
works, the ad hoc prescription of Landau and Lifshitz is unsatisfactory.

To get sensible solutions to the Lorentz-Dirac equation, its new degree
of freedom must be continuously fine-tuned so that the trajectory
stays on a submanifold of the phase space (the critical submanifold).
Spohn used modern ideas in the theory of dynamical systems to determine
the equation of motion projected to this critical submanifold: it
is precisely the above Landau-Lifshitz equation! Thus we consider
the century-old problem of determining the equation of motion of a
radiating point charge to be settled finally. 

However, the electron has a magnetic moment as well as electric charge;
radiation arises not only from the acceleration of the charge but
also the precession of the magnetic moment. In the relativistic case,
these are of comparable magnitude. Therefore, it is necessary to extend
the Lorentz-Dirac and Landau-Lifshitz equations to the case of a spinning
charged particle. 

Bargman, Michel and Telegdi\cite{BMT } proposed, more than a half
century ago, an equation for a (non-radiating) classical particle
with charge and magnetic moment. However, it does not seem to be the
classical limit of the Dirac wave equation of the electron. The correct
equation must have a canonical form whose quantization leads to the
Dirac wave equation. The BMT equation does not appear to satisfy this
condition.

A better approach was proposed more recently by Barut and Zanghi\cite{BZ}.
We will review their ideas, which show that a canonical quantization
does lead to the Dirac equation. We will also find solutions of the
Barut-Zanghi equation for a non-radiating charged particle in a constant
magnetic field: itself a new result. 

Barut and Unal\cite{BU} also derived the analogue of the Lorentz-Dirac
equation, after a mass renormalization. However, it suffers from the
same runaway solutions of the free particle, as the spinless case.

In this paper, we propose an equation of motion for a radiating spinning
charged particle with gyromagnetic ratio $2$ (e.g., the electron
or the muon) by applying a prescription analogous to that of Landau
and Lifshitz to the Barut-Zanghi equation. We will show that, at least
for spin-polarized orbits in a  constant magnetic field, the orbits decay as expected physically.
It should be possible to test our equations using a small (few MeV)
electron synchrotron or a muon storage ring, such as the one used
for the determination of $g-2$. Also, it should not be difficult
to revise the predictions for an electron accelerated by a laser field\cite{LaserLL}.
Current predictions do not include the effect of the magnetic moment.

We do not attempt, in this paper, to derive our prescription from
first principles as Spohn did.

\section{Non-Radiating Spinning Charged Particle}

Can we talk of the motion of a particle with spin in the classical
limit? At first this would appear to be impossible. But we do it all
the time. The motion of electrons in synchrotron of a radius of several
meters (if not kilometers) is surely classical: even if the spin were
polarized, the differences in the quantum energy levels of the electron
are much too small to be of significance. If we can find a lagrangian
whose canonical quantization is the Dirac wave equation, it can be
the basis for such a classical treatment.

The lagrangian proposed by Barut and Zanghi\footnote{They use a slightly different system of units.} is,

\beq
L=\frac{i}{2}\left[\dot{\bar{z}}z-\bar{z}\dot{z}\right]+p_{\mu}\dot{x}^{\mu}-H,\quad H=\bar{z}\gamma^{\nu}z\left[p_{\nu}-qA_{\nu}\right],\,\,\,\,\,\bar{z}=z^{\dagger}\gamma_{0}
\eeq 

Here $z$ is a Dirac spinor with four complex (not Grassmann) numbers
as components. The dimensions of $z$ are that of the square root
of action (=mass{*}length) and that of $\tau$ is the inverse of mass,
in units where $c=1$ but $\hbar\neq1.$ Then $qF_{\mu\nu}$ has dimensions
of mass over length. 

The equations of motion are, with $\pi_{\mu}=p_{\mu}-qA_{\mu}$ 

\begin{eqnarray}
\frac{dx^{\mu}}{d\tau}&=&\bar{z}\gamma^{\mu}z \nonumber \\ 
-i \frac{dz}{d\tau}&=&\gamma^{\mu}\pi_{\mu}z  \nonumber \\ 
\frac{d\pi_{\mu}}{d\tau}&=&qF_{\mu\nu}\bar{z}\gamma^{\nu}z \nonumber \\
\end{eqnarray}

Note that, $\tau$ is not proper time as  $\dot{x}^{\mu}$
is not  necessarily of constant length. But we have the orthogonality
condition 

\beq
\dot{x}^{\mu}\dot{\pi}_{\mu}=0.
\eeq

These equations follow from the Hamiltonian 
\beq
H=\bar{z}\gamma^{\mu}z\pi_{\mu}
\eeq

and Poisson Brackets (P.B.) are obtained using 

\beq
\{f,g\}=i\{\frac{\partial f}{\partial z}\frac{\partial g}{\partial\bar{z}}-\frac{\partial g}{\partial z}\frac{\partial f}{\partial\bar{z}}\}+g_{\mu\nu}\{\frac{\partial f}{\partial p^{\mu}}\frac{\partial g}{\partial x^{\nu}}-\frac{\partial f}{\partial x^{\mu}}\frac{\partial g}{\partial p^{\nu}}\}
\eeq

\beq
\left\{ \pi_{\mu},\pi_{\nu}\right\} =-qF_{\mu\nu},\quad\left\{ \pi_{\mu},x^{\nu}\right\} =\delta_{\mu}^{\nu} \nonumber
\eeq

\beq
\left\{ z,\bar{z}\right\} =i,\quad\left\{ \pi_{\mu},z\right\} =0=\left\{ z,x\right\} 
\eeq

All other pairs of P.B. are zero. We can introduce a constraint
\beq
\bar{z}z=a
\eeq

consistent with the P.B. and the equations of motion. The quantity
$a$, with the dimensions of action, sets the scale for the intrinsic
angular momentum of the particle.

Upon quantization, the wave function can be thought of as a function
of $x$, $p_{\mu}=\frac{\hbar}{i}\frac{\partial}{\partial x^{\mu}}$;
it is also a polynomial in $\bar{z}$, with $z=-\hbar\frac{\partial}{\partial\bar{z}}$
satisfying the constraint 
\beq
-\hbar\bar{z}\frac{\partial}{\partial\bar{z}}\psi=a\psi.
\eeq

Thus the parameter $a$ is quantized in multiples of $\hbar$. The
smallest value is $a=\hbar$ for which $\psi(x,\bar{z})=\bar{z}\psi(x)$
is a linear function.

The Dirac wave equation is the eigenvalue equation for $H$,
with eigenvalue  $am$:

\beq
i\hbar\gamma^{\mu}\frac{\partial}{\partial x^{\mu}}\psi=m\psi
\eeq

Other quantizations, for which$\psi(\bar{z})$ is a higher degree
polynomial, correspond to the wave equations of Bhabha and Harish-Chandra.
They could be interesting as wave equations for composite particles
like hadrons\cite{SkyrmeWaveEqn}.

It is convenient to introduce the velocity and spin variables

\beq
v^{\mu}=\bar{z}\gamma^{\mu}z,\quad S^{\mu\nu}=\frac{i}{4}\bar{z}[\gamma^{\mu},\gamma^{\nu}]z
\eeq

so that 

\begin{eqnarray}
\frac{dx^{\mu}}{d\tau}&=&v^{\mu} \nonumber \\
\frac{dv^{\mu}}{d\tau}&=&4S^{\mu\nu}\pi_{\nu} \nonumber\\
\frac{d\pi_{\mu}}{d\tau}&=&qF_{\mu\nu}v^{\nu} \nonumber \\
\frac{dS^{\mu\nu}}{d\tau}&=&v^{\nu}\pi^{\mu}-v^{\mu}\pi^{\nu} \nonumber\\
\end{eqnarray}

These equations describe a relativistic analogue of the isotropic
precessing top of Poinsot \cite{LandauLifshitzMech}, as we explain  in the Appendix.
The hamiltonian
\beq
H=v^{\mu}\pi_{\mu}
\eeq

 describes a coupling to the external field which causes
a precession of the velocity and spin. For a constant electromagnetic
field, $K=\pi\cdot\pi-qS^{\mu\nu}F_{\mu\nu}$ is conserved as well:
\beq
\frac{d}{d\tau}\left[\pi\cdot\pi-qS^{\mu\nu}F_{\mu\nu}\right]=2qF_{\mu\nu}\pi^{\mu}v^{\nu}-2qF_{\mu\nu}v^{\nu}\pi^{\mu}=0.
\eeq

This is the classical analogue of the square of the Dirac operator,
corresponding to the gyromagnetic ratio $2.$ We will see later that
it is proportional to the angular momentum in the plane of the electromagnetic
field.

\section{A Non-Radiating Spinning Particle in a Constant Magnetic Field}

Now let us turn to solving the equations of motion in the case of
a magnetic field pointed along the third direction: $F_{12}=-F_{21}>0$;
and all other components are zero. Before we solve the problem including
radiation damping, we must recover the solution without radiation.
Even this appears to be new: Barut and collaborators only solved the
case of a free particle. 

We consider only orbits that lie in the $12$ plane. That is, we restrict
to the case where all the tensor indices $\mu,\nu,\rho\cdots$ are
in the $012$ subspace. It is useful to introduce the `dual' variables

\begin{eqnarray}
q F_{\mu\nu}&=&\epsilon_{\mu\nu\rho}F^{\rho} \nonumber\\
S^{\mu\nu}&=&\epsilon^{\mu\nu\rho}S_{\rho} \nonumber \\
S_{\rho}&=&\frac{1}{2}\epsilon^{\mu\nu\rho}S_{\mu\nu} \nonumber\\
\end{eqnarray}

For a constant magnetic field, only the time component $F^{0}$ is
non-zero. We can assume it to be positive without loss of generality.

Then

\begin{eqnarray}
\frac{d\pi_{\mu}}{d\tau}&=&\epsilon_{\mu\nu\rho}v^{\nu}F^{\rho} \nonumber\\
\frac{dv^{\mu}}{d\tau}&=&-4\epsilon^{\mu\nu\rho}S_{\nu}\pi_{\rho} \nonumber\\
\frac{dS_{\mu}}{d\tau}&=&-\epsilon_{\mu\nu\rho}v^{\nu}\pi^{\rho} \nonumber\\
\end{eqnarray}

If we define \[
\lambda_{\mu}=\frac{1}{2}(S_{\mu}+\frac{1}{2}v_{\mu}),\quad\rho_{\mu}=\frac{1}{2}(S_{\mu}-\frac{1}{2}v_{\mu})\]

they reduce to 

\begin{eqnarray}
\frac{d\pi_{\mu}}{d\tau}&=&2\epsilon_{\mu\nu\rho}\left[\lambda^{\nu}-\rho^{\nu}\right]F^{\rho} \nonumber \\
\frac{d\lambda^{\mu}}{d\tau}&=&-2\epsilon^{\mu\nu\rho}\lambda_{\nu}\pi_{\rho} \nonumber \\
\frac{d\rho_{\mu}}{d\tau}&=&2\epsilon_{\mu\nu\rho}\rho^{\nu}\pi^{\rho} \nonumber \\
\end{eqnarray}

\subsection{Spin-Polarized Orbits}

There is a special class of `left-handed' orbits for which $\rho_{\mu}=0$.
(Also, the opposite case with $\lambda_{\mu}=0$.) For these, the
spin and velocity are synchronized in the following way:

\beq
S_{\mu\nu}=\frac{1}{2}\epsilon_{\mu\nu\rho}v^{\rho}
\eeq

In the non-relativistic limit, this means the spin is pointed along
the magnetic field (times $q$).

Since these solutions are periodic (not chaotic), orbits of this kind
in a synchrotron should be of special interest. If the gyromagnetic
ratio is slightly different from two, the spin would depart from this
orientation, which is the principle behind a famous experiment for
the measurement of $g-2$ for the muon.

The equations reduce to 

\begin{eqnarray}
\frac{d\pi_{\mu}}{d\tau}&=&\epsilon_{\mu\nu\rho}v^{\nu}F^{\rho} \nonumber \\
\frac{dv^{\mu}}{d\tau}&=&-2\epsilon^{\mu\nu\rho}v_{\nu}\pi_{\rho} \nonumber \\
\end{eqnarray}

The tensor $\epsilon_{\mu\nu}^{\sigma}=\epsilon_{\mu\nu\rho}\eta^{\rho\sigma}$
form the structure constants of the three dimensional Lorentz Lie
algebra $SO(1,2)$. This is analogous to the way that the Levi-Civita
tensor gives the stucture constants of the Lie algebra of rotations
in Euclidean three dimensional space. Denote the Lie product of $SO(1,2)$
as a relativistic cross product

\beq
\left(u\times w\right)_{\mu}=\epsilon_{\mu}^{\cdot\nu\rho}u_{\nu}w_{\rho}
\eeq

This differs from the familiar cross product in Euclidean geometry
by some signs:

\beq
\left(u\times w\right)_{0}=u_{1}w_{2}-u_{2}w_{1},\quad\left(u\times w\right)_{1}=-u_{2}w_{0}+u_{0}w_{2},\quad\left(u\times w\right)_{2}=-u_{0}w_{1}+u_{1}w_{0}
\eeq

Then our equations are a relativistic version of the equation for
a precessing top:

\beq
\dot{\pi}=v\times F,\quad\dot{v}=-2v\times\pi
\eeq

It follows that 

\beq
v\cdot\dot{v}=0
\eeq

so that $v\cdot v=a^{2}$ is conserved; $a$ is some number (like
Planck's constant) with the dimensions of action; the spin is of magnitude
$\frac{a}{2}.$ Also, the parameter $\tau$ is  proportional to proper time in this special case.

The energy  $\pi_{0}=E$ and 

\beq
K=\pi_{\mu}\pi^{\mu}-v^{\mu}F_{\mu}
\eeq
are conserved\footnote{$K$ is proportional  to angular momentum.%
}. Now,

\beq
\dot{v}_{0}=2(v_{2}\pi_{1}-v_{1}\pi_{2})
\eeq
and

\beq
\dot{v}_{0}^{2}=4\left(\left[v_{1}^{2}+v_{2}^{2}\right]\left[\pi_{1}^{2}+\pi_{2}^{2}\right]-\left[v_{1}\pi_{1}+v_{2}\pi_{2}\right]^{2}\right)
\eeq

Re-expressing the r.h.s. in terms of conserved quantities,

\beq
\dot{v}_{0}=2\sqrt{\left[v_{0}^{2}-a^{2}\right]\left[E^{2}-K+v_{0}F_{0}\right]-\left[H-Ev_{0}\right]^{2}}
\eeq

Weierstrass elliptic function satisfy this differential equation.
The remaining components of $\pi,v$ can be obtained by solving the
other constraint equations. It is also possible to understand the
solution as an integrable hamiltonian system, a relativistic analogue
of the precessing top. (See the Appendix)

\section{The Equation of Motion for Radiating Spinning Particles}

Barut and Unal followed the method of Dirac to derive the analogue
of the Lorentz-Dirac equation of motion including the self-force.
Only the equation for $\pi_{\mu}$ is affected:

\beq
\frac{d\pi_{\mu}}{d\tau}=qF_{\mu\nu}v^{\nu}+q^2\left[\eta_{\mu\nu}-\frac{v_{\mu}v_{\nu}}{v\cdot v}\right]\left\{ \frac{2}{3}\frac{\ddot{v}^{\nu}}{v\cdot v}-\frac{9}{4}\frac{v\cdot\dot{v}\dot{v}^{\nu}}{\left(v\cdot v\right)^{2}}\right\} 
\eeq

The last term is the contribution from self-interactions, after a
renormalization. It is a singular perturbation i.e. $q^{2}$
is small and it changes the order of the equations. It has runaway
solutions, just like the Lorentz-Dirac equation. Let us apply to it
a prescription like that of  Landau and Lifshitz, to get a second
order equation of motion. 

Naively, the Landau-Lifshitz prescription amounts to using 

\beq
\ddot{v}^{\mu}=4\dot{S}^{\mu\nu}\pi_{\nu}+4S^{\mu\nu}\dot{\pi_{\nu}}
\eeq 
and replacing $\dot{\pi}$ by its zeroth order contribution:

\beq
\ddot{v}^{\mu}\to4\dot{S}^{\mu\nu}\pi_{\nu}+4S^{\mu\nu}qF_{\nu\rho}v^{\rho}=4\left[v^{\nu}\pi^{\mu}-v^{\mu}\pi^{\nu}\right]\pi_{\nu}+4S^{\mu\nu}qF_{\nu\rho}v^{\rho}
\eeq

But this can't be right: for a free particle ($F_{\mu\nu}=0$), the
radiation reaction force should be zero. Dropping the terms not proportional
to $F_{\mu\nu}$, 

\beq
\frac{d\pi_{\mu}}{d\tau}=qF_{\mu\nu}v^{\nu}+\frac{8q^{3}}{3v\cdot v}\left[\eta_{\mu\nu}-\frac{v_{\mu}v_{\nu}}{v\cdot v}\right]S^{\nu\rho}F_{\rho\sigma}v^{\sigma}
\eeq

Along with the previous equations for $v$ and $S,$ this system of
ordinary differential equations is our proposal for the equation of
motion of a spinning radiating charged particle. We must verify that
the solutions are physically sensible: there should not be runaway
solutions. We will study the case of polarized orbits in a constant
magnetic field and show that the energy decreases with time. The ultimate
test of our proposal must be experimental, especially since we do
not yet have a derivation from first principles.

\section{Radiating Spin-Polarized Particles in a Magnetic Field}

Now let us turn to our proposed equations for a radiating spinning
charged particle. Again, we will look for solutions  in the special
case where the spin is polarized 

\beq
S_{\mu}=\frac{1}{2}v_{\mu}
\eeq

Since the radiative terms do not change the equations for $v^{\mu},\, S^{\mu\nu}$
this condition is still preserved by time evolution i.e. $\dot{S}^{\mu}=\frac{1}{2}\dot{v}^{\mu}$.
Unpolarized orbits are chaotic and the radiation damping is likely
to be even larger. (See Appendix)

The equations then reduce to 

\begin{eqnarray}
\frac{d\pi_{\mu}}{d\tau}&=&qF_{\mu\nu}v^{\nu}+\frac{4q^{3}}{3v\cdot v}\left[\eta_{\mu\nu}-\frac{v_{\mu}v_{\nu}}{v\cdot v}\right]\epsilon^{\nu\rho\alpha}v_{\alpha}F_{\rho\sigma}v^{\sigma} \nonumber \\
\frac{dv^{\mu}}{d\tau}&=&-2\epsilon^{\mu\nu\rho}v_{\nu}\pi_{\rho} \nonumber \\
\end{eqnarray}

We know that, $v\cdot\dot{v}=0$ and $v\cdot v=v_{0}^{2}-(v_{1}^{2}+v_{2}^{2})=a^{2}.$ 

\beq
\frac{d\pi_{\mu}}{d\tau}=\epsilon_{\mu\nu\rho}v^{\nu}F^{\rho}+\left[a^{2}\eta_{\mu\nu}-v_{\mu}v_{\nu}\right]\gamma F^{\nu}
\eeq

where 

\beq
\gamma=\frac{4q^{2}}{3a^{2}}
\eeq

In terms of relativistic cross products

\beq
\dot{\pi}=v\times[F+\gamma F\times v],\quad\dot{v}=-2v\times\pi
\eeq

Thus, even the radiation damping terms have a simple meaning in terms
of the relativistic cross product of the Lie algebra $SO(1,2).$

Note that 
\beq
H=v\cdot\pi
\eeq

is still conserved and $m=\frac{H}{a}$ continues to have the physical
meaning of mass. But energy $E=\pi_{0}$ and $K$ are no longer conserved.
The evolution of $v_{0}$ continues to be determined by the equation
we derived earlier. Thus we get the system 

\begin{eqnarray}
\frac{dE}{d\tau}&=&-\gamma F_{0}(v_{0}^{2}-a^{2}) \nonumber \\
\frac{dK}{d\tau}&=&2\gamma F_{0}\left[a^{2}E-Hv_{0}\right] \nonumber \\
\dot{v}_{0}&=&2\sqrt{\left[v_{0}^{2}-a^{2}\right]\left[E^{2}-K+v_{0}F_{0}\right]-\left[H-Ev_{0}\right]^{2}} \nonumber \\
\end{eqnarray}

Since $E$ and $K$ are no longer constants, the solution is no longer
an elliptic function. Also, noting that $F_{0}>0$ in our convention
and $v_{0}^{2}-a^{2}=(v_{1}^{2}+v_{2}^{2})>0$, we see that energy
is monotonically decreasing.

\section*{Acknowledgement}

We thank Gregory Bentsen, V. Sreedhar and Tamar Friedman for discussions.
This work was supported in part by a grant from the US Department
of Energy under contract DE-FG02-91ER40685.

\section*{Appendix: Hamiltonian Formalism}

The Barut-Zanghi equations for a spinning charges particle follow
from the P. B. forming the Lie algebra $SO(2,3)$:

\begin{eqnarray}
\left\{ v^{\mu},v^{\nu}\right\} &=& 4S^{\mu\nu} \nonumber \\
\left\{ S^{\mu\nu},v^{\rho}\right\} &=& \eta^{\mu\rho}v^{\nu}-\eta^{\nu\rho}v^{\mu} \nonumber \\
\left\{ S^{\mu\nu},S^{\rho\sigma}\right\} &=& \eta^{\mu\rho}S^{\nu\sigma}+\eta^{\nu\sigma}S^{\mu\rho}-\eta^{\nu\rho}S^{\mu\sigma}-\eta^{\mu\sigma}S^{\nu\rho} \nonumber \\
\end{eqnarray}

Using previous relations we can also find the following P.B. : 

\begin{eqnarray}
\left\{ v_{\mu},v_{\nu}\right\} &=& 4S_{\mu\nu} \nonumber \\
\left\{ S_{\mu},S_{\nu}\right\} &=& S_{\mu\nu} \nonumber \\
\left\{ S_{\mu},v_{\nu}\right\} &=& \epsilon_{\mu\nu\sigma}v^{\sigma} \nonumber \\
\end{eqnarray}

which generalizes the rotation algebra. When the third axis is ignored,
this reduces to $SO(2,2)\approx SO(1,2)\oplus SO(1,2)$. This splitting
can be seen by in terms of the linear combinations $\lambda_{\mu},\rho_{\mu}:$

\begin{eqnarray}
\left\{ \lambda_{\mu},\lambda_{\nu}\right\} &=& \epsilon_{\mu\nu\sigma}\lambda^{\sigma} \nonumber \\
\left\{ \rho_{\mu},\rho_{\nu}\right\} &=& \epsilon_{\mu\nu\sigma}\rho^{\sigma} \nonumber \\
\left\{ \lambda_{\mu},\rho_{\nu}\right\} &=& 0 \nonumber \\
\end{eqnarray}

We have also the P.B., 

\begin{eqnarray}
\left\{ \lambda_{\mu},\pi_{\nu}\right\} &=& 0 \nonumber \\
\left\{ \rho_{\mu},\pi_{\nu}\right\} &=& 0 \nonumber\\  
\left\{ \pi_{\mu},\pi_{\nu}\right\} &=& -\epsilon_{\mu\nu\rho}F^{\rho} \nonumber \\
\end{eqnarray}

The hamiltonian is then

\beq
H=2[\lambda-\rho]\cdot\pi
\eeq

\subsection*{Non-Radiating Spin-Polarized Orbits in Constant Magnetic Field}

Since

\beq
v^{0}=\sqrt{a^{2}+v_{b}v_{b}}
\eeq

and $\pi_{0}=E$ is conserved we only have four co-ordinates in the
phase space: $\pi_{a},v_{a}$. The P.B. are 

\beq
\left\{ \pi_{1},\pi_{2}\right\} =-F_{0}
\eeq

and 
\beq
\left\{ v_{1},v_{2}\right\} =2\sqrt{a^{2}+v_{1}^{2}+v_{2}^{2}}
\eeq

The latter corresponds to  the standard symplectic form on hyperbolic space. It is not hard
to find canonical co-ordinates

$$\{\frac{\pi_{1}}{\sqrt{F_{0}}},\frac{\pi_{2}}{\sqrt{F_{0}}}\}\rightarrow\{\sqrt{2p}\cos\theta,\sqrt{2p}\sin\theta\},\\
\{v_{1},v_{2}\}\rightarrow\{a\,\sinh\theta\,\cos\phi,a\,\sinh\theta\,\sin\phi\}$$

so that-

\beq
\left\{ p,\theta\right\} =1,\quad p=\frac{1}{2F_{0}}(\pi_{1}^{2}+\pi_{2}^{2}),\quad\theta=\arctan\frac{\pi_{2}}{\pi_{1}}
\eeq

\beq
\left\{ v_{0},\phi\right\} =1,\quad\phi=\arctan\frac{v_{2}}{v_{1}}
\eeq

Note that $J=p+\frac{1}{2}v_{0}=\frac{K}{2F_{0}}$ is the generator
of rotations; it is a conserved quantity corresponding to angular
momentum. The hamiltonian is then

\beq
H=Ev_{0}-\sqrt{2F_{0}\left[v_{0}^{2}-a^{2}\right](J-\frac{1}{2}v_{0})}\cos\left(\theta-\phi\right)
\eeq

Since

\beq
pd\theta+\frac{1}{2}v_{0}d\phi=Jd\theta+v_{0}d\alpha,\quad\alpha=\frac{1}{2}(\phi-\theta)
\eeq

we can use the pairs $J,\theta$ and $v_{0},\alpha$ as canonical
variables:

\beq
H=Ev_{0}-\sqrt{2F_{0}\left[v_{0}^{2}-a^{2}\right](J-\frac{1}{2}v_{0})}\cos2\alpha
\eeq

The equation of motion is

\beq
\frac{dv_{0}}{d\tau}=-\frac{\partial H}{\partial\alpha}=-2\sqrt{F_{0}\left[v_{0}^{2}-a^{2}\right](2J-v_{0})}\sin2\alpha
\eeq

Thus

\beq
\left[\frac{dv_{0}}{d\tau}\right]^{2}=4F_{0}\left[v_{0}^{2}-a^{2}\right](2J-v_{0})-4\left(H-Ev_{0}\right)^{2}
\eeq

which is the Weierstrass equation. Because of its polarization, this
also determines the spin.

Also

\beq
\frac{d\theta}{d\tau}=\frac{\partial H}{\partial J}=-\sqrt{\frac{F_{0}\left[v_{0}^{2}-a^{2}\right]}{2J-v_{0}}}\cos2\alpha
\eeq

Thus

\beq
\frac{d\theta}{d\tau}=\frac{H-Ev_{0}}{(2J-v_{0})}
\eeq

which determines $\theta$ also in terms of elliptic integrals. 

To determine the position variable, we note the that 

\beq
\frac{d}{d\tau}\left[\pi_{\mu}-\epsilon_{\mu\nu\rho}x^{\nu}F^{\rho}\right]=0
\eeq

This leads to the conserved quantity (corresponding to translation
invariance)

\beq
X_{\mu}=\pi_{\mu}-\epsilon_{\mu\nu\rho}x^{\nu}F^{\rho}
\eeq

Thus (in a co-ordinate system in which $F_{0}>0,F_{a}=0$ and $X_{c}=0$)
we have 

\beq
x_{b}=-\frac{1}{F_{0}}\epsilon_{bc}\pi_{c}
\eeq

In other words, knowing $\pi_{1},\pi_{2}$ is the same knowing the
spatial orbit, except for a rotation by $90^{\circ}$ and a scaling
by $F_{0}.$ In polar co-ordinates this amounts to knowing $p=J-\frac{1}{2}v_{0}$
and $\theta$ as functions of $\tau$, a kind of elliptic curve:

\beq
\left[\frac{dp}{d\tau}\right]^{2}=2F_{0}\left[4(J-p)^{2}-a^{2}\right]p-\left(H-2EJ+2Ep\right)^{2},\quad\frac{d\theta}{d\tau}=\frac{H-EJ}{2p}+\frac{E}{2}
\eeq

It is not correct to interpret the $x$ as the position of the electron:
due to the phenomenon of zitterbewegung. Even for a free particle,
$x^{\mu}(\tau)$ is a spiral and not a straightline. Barut et. al.
have shown how to recover a center of mass variable for the electron
from this complicated solution.

\subsection*{A Symplectic Reduction of the General Planar Orbit}

The general case of planar but not spin-polarized orbits does not
appear to be integrable. It corresponds to chaotic motion of an electron
in a synchrotron. We can use the conserved quantities 

\beq
\pi_{0}=E,\ \lambda\cdot\lambda\equiv a^{2},\ \rho\cdot\rho\equiv b^{2} 
\eeq

to reduce it to a conservative system with three degrees of freedom:
the phase space is a plane ($\pi_{a}$) times a pair of hyperboloids
parametrized by ($\lambda_{\mu},\rho_{\mu}$). It is easy to check
that 

\beq
\left\{ p,\theta\right\} =1
\eeq

with
\beq
p=\frac{1}{2F_{0}}\pi_{a}\pi_{a},\ \theta=\arctan\frac{\pi_{2}}{\pi_{1}}
\eeq

Also,

\begin{eqnarray}
\left\{ \lambda_{0},\phi\right\} =1,\quad\phi=\arctan\frac{\lambda_{2}}{\lambda_{1}} \nonumber \\
\left\{ \rho_{0},\chi\right\} =1,\quad\chi=\arctan\frac{\rho_{2}}{\rho_{1}} \nonumber \\
\end{eqnarray} 

In terms of these 

\begin{eqnarray}
H&=&2E(\lambda_{0}-\rho_{0})-2(\lambda_{a}-\rho_{a})\pi_{a} \nonumber \\
&=&E(\lambda_{0}-\rho_{0})-\sqrt{2F_{0}\left[\lambda_{0}^{2}-a^{2}\right]p}\cos[\phi-\theta]+\sqrt{2F_{0}\left[\rho_{0}^{2}-b^{2}\right]p}\cos[\chi-\theta] \nonumber \\
\end{eqnarray}

Another conserved quantity is the generator of rotations%
\footnote{This is linearly related to the quantity $K$ we introduced earlier:
\[
K=2F_{0}J\]
}

\beq
J=p+\lambda_{0}+\rho_{0}
\eeq

The reduced system has three degrees of freedom and two conserved
quantities $H,J$ : just one conserved quantity short of an integrable
system. When $\lambda_{0}=a$ (or $\rho_{0}=b$ ) one degree of freedom
decouples and we get an integrable system. Circular orbits correspond
to the case $\phi=\chi,\lambda_{0}+\rho_{0}=\mathrm{constant}.$

It may be more convenient to use $\alpha=\phi-\theta,\beta=\chi-\theta,\theta$
as the independent position variables. Then 

\beq
p_{\alpha}d\alpha+p_{\beta}d\beta+p_{\theta}d\theta=pd\theta+\lambda_{0}d\phi+\rho_{0}d\chi=pd\theta+\lambda_{0}d[\theta+\alpha]+\rho_{0}d[\theta+\beta]
\eeq

\begin{eqnarray}
p_{\alpha}&=&\lambda_{0} \nonumber \\
p_{\beta} &=& \rho_{0} \nonumber \\
p_{\theta} &=& p+\lambda_{0}+\rho_{0}=J \nonumber \\ 
\end{eqnarray}

Then we get a system with two degrees of freedom with hamiltonian
\beq
H=E(\lambda_{0}-\rho_{0})-\sqrt{2F_{0}\left[\lambda_{0}^{2}-a^{2}\right][J-\lambda_{0}-\rho_{0}]}\cos\alpha+\sqrt{2F_{0}\left[\rho_{0}^{2}-b^{2}\right][J-\lambda_{0}-\rho_{0}]}\cos\beta
\eeq

This appears to be a chaotic system. In general, the information
about the polarization of spin in the initial condition will be lost
rapidly and the particle beam will look unpolarized. For the special
case $\rho_{0}=0=b$ the spin remains polarized and solutions can
be obtained in terms of elliptic functions as we saw above.

\end{document}